\documentclass[aps, preprint, showkeys, nofootinbib]{revtex4}%
\usepackage{amsmath}
\usepackage{amsfonts}
\usepackage{amssymb}
\usepackage{graphicx}%
\providecommand{\U}[1]{\protect\rule{.1in}{.1in}}
\begin{document}

\preprint{ }
\title{A brief note on the limit $\omega\rightarrow\infty$ in Weyl geometrical scalar-tensor theory}
\author{A. Barros}
\affiliation{Centro de Desenvolvimento Sustent\'{a}vel do Semi\'{a}rido, Universidade
Federal de Campina Grande, 58540-000, Sum\'{e}, PB, Brazil.\\atbarros@ufcg.edu.br}
\author{C. Romero}
\affiliation{Departamento de F\'{\i}sica, Universidade Federal da Para\'{\i}ba. C.P. 5008,
58059-970, Jo\~{a}o Pessoa, PB, Brazil.\\cromero@fisica.ufpb.br}

\begin{abstract}
We obtain vacuum solutions in the presence of a cosmological constant in the
context of the Weyl geometrical scalar-tensor theory. We investigate the limit
when $\omega$ goes to infinity and show by working out the solutions that in
this limit there are some cases in which the scalar field tends to a constant
(with the implicit consequence of \ the geometry becoming Riemannian),
although the solutions do not reduce to the corresponding Einstein solutions.
We have also extended a previous result, known in the literature, by showing
that in the case of vacuum with cosmological constant the field equations of
the Weyl geometrical scalar-tensor theory are formally identical to Brans-Dicke field equations, even though these theories are not physically
equivalent. 

\end{abstract}


\maketitle

\section{Introduction}
Among the so-called modified gravity theories one\ that stands as the most
simple and popular is that proposed by C. Brans and R. Dicke in
1961\cite{Bra61}. As is well known, the main motivation of the authors was to
incorporate Mach's principle into a relativistic theory of gravity. With this
purpose\ in mind they managed to formulate a scalar-tensor theory, in which
the gravitational effects would be described both by a metric field $g_{\mu
\nu}$ and a scalar field $\Phi$, with the geometry of the underlying
space-time manifold being assumed to be Riemannian. The scalar field, which is
neither a geometrical nor a matter-related field, replaces the gravitational
constant and is interpreted as the inverse of the gravitational coupling
parameter. To date Brans-Dicke theory which is generally considered to be in
agreement with observation, has been regarded with interest by many
theoreticians\cite{Will}.

A different approach to a scalar-tensor theory of gravity consists in
considering the scalar field $\Phi$ to be of a geometrical nature, which would
be more in accordance with the geometrization program of physics that Einstein
started with general relativity. In this sense, the scalar field which appears
in the so-called Weyl geometrical scalar-tensor theory is regarded as an
essential element of the space-time geometry\cite{rom14}. Indeed, in this
theory space-time is assumed to be a very special case of Weyl (non-integral)
space-time\cite{wey18}. This has been referred to in the literature as a
\textit{Weyl integrable space-time }(WIST)\cite{rom08}. Other gravity theories
in which a scalar field plays a geometrical role have been
proposed\cite{Fonseca}.

Due to some similarities between the Weyl geometrical scalar-tensor theory and
the Brans-Dicke theory, it is interesting to see what happens when $\omega$,
the scalar field coupling constant, goes to infinity. Concerning this point,
let us firstly recall that for\ quite a long time it was (erroneously) held
that Brans-Dicke theory would reduce to general relativity when $\omega
\rightarrow\infty$\cite{Weinberg}. The question now seems to be settled after
the publication of a series of articles on the subject\cite{Faraoni}. In fact,
it has been shown that under certain circumstances Brans-Dicke theory does not
go over general relativity when $\omega\rightarrow\infty$. This result is not
entirely inconsequential as it may affect the way we set limits on the values
of $\omega$ from solar system experiments or even from cosmological
observation. In this paper, we examine the same kind of asymptotic behaviour,
however now in the context of another scalar-tensor theory which also contains
an adimensional parameter, namely, Weyl geometrical scalar-tensor theory.

Inspired by previous work\cite{bar93}, our approach to this problem will
consist in initially obtaining some solutions of the Weyl geometrical
scalar-tensor theory, considering the vacuum field equations in the presence
of a cosmological term $\Lambda$. With the help of these solutions, we shall
study the limit $\omega\rightarrow\infty$, investigating the possibility of
obtaining the general relativity equations.

Finally, a relevant question is concerned with the physical motivation of the
present work. As a matter of fact, the problem we have examined in the present
paper is of purely theoretical nature: We investigate a mathematical feature
of a class of scalar-tensor theories which have a free parameter $\omega$. The
behaviour of different solutions when $\omega$ goes over infinity may have
consequences in the interpretation of the theory. For instance, it was long
believed that the original Brans-Dicke theory would approach general
relativity for large $\omega$, in which case the theory would completely lose
its interest since on the basis of the Occam's rasor principle the simplest
theory is always preferable. \ Now it is well known that the limit of
Brans-Dicke solutions to general relativistic solutions corresponding to the
same momentum-energy tensor is not unique, or may not yield a solution of
Einstein's equations at all, in which case they must not simply be discarded
as devoid of interest even for large $\omega$. For additional motivation
concerning this question, which apply both to the original Brans-Dicke theory and the geometrical scalar-tensor theory, we would like to refer the reader to Ref. 10.

The paper is organized as follows. In Section 2, we show that if one considers
the vacuum field equations with cosmological constant, the field equations of
the Weyl geometrical scalar-tensor theory and Brans-Dicke theory are formally
identical. Based on this result, we exhibit solutions of the Weyl geometrical
scalar-tensor theory in Section 3. We proceed in Section 4 to examine the
limit $\omega\rightarrow\infty$ for some cases. Finally, Section 5 is devoted
to our conclusions.

\section{The Weyl geometrical scalar-tensor theory}
The field equations of the Weyl geometrical scalar-tensor theory in its most
general form are given by\cite{rom16}
\begin{equation}
\bar{G}_{\mu\nu}=\omega(\phi)\left(  \frac{\phi_{,\alpha}\phi^{,\alpha}}%
{2}g_{\mu\nu}-\phi_{,\mu}\phi_{,\nu}\right)  -\frac{1}{2}e^{\phi}g_{\mu\nu
}V(\phi)-8\pi T_{\mu\nu}%
\end{equation}%
\begin{equation}
\bar{\square}\phi=-\left(  1+\frac{1}{2\omega}\frac{d\omega}{d\phi}\right)
\phi_{,\mu}\phi^{,\mu}-\frac{e^{\phi}}{\omega}\left(  \frac{1}{2}\frac
{dV}{d\phi}+V\right)  ,
\end{equation}
where $\phi$ is a scalar field, $\omega$ is a function of $\phi$, $V(\phi)$
corresponds to the scalar field potential and $T_{\mu\nu}$ represents the Weyl
invariant energy-momentum tensor of the matter fields.\cite{rom14} Moreover,
we are denoting by $\bar{G}_{\mu\nu}$ and $\bar{\square}$ the Einstein tensor
and the d'Alembertian operator, respectively, as defined with respect to the
Weyl connection, whose coefficients in a local coordinate basis read
\begin{equation}
\Gamma_{\mu\nu}^{\alpha}=\{_{\mu\nu}^{\alpha}\}-\frac{1}{2}g^{\alpha\beta
}(g_{\beta\mu}\phi_{,\nu}+g_{\beta\nu}\phi_{,\mu}-g_{\mu\nu}\phi_{,\beta}),
\end{equation}
with $\{_{\mu\nu}^{\alpha}\}$ representing the usual Christoffel symbols. At
this point, we should mention that the scalar field $\phi$ is regarded as a
purely geometrical field. Indeed, $\phi$ is a basic ingredient essential of
the Weyl nonmetricity condition
\begin{equation}
\triangledown_{\alpha}g_{\mu\nu}=g_{\mu\nu}\phi_{,\alpha}%
\end{equation}
which in this form characterizes the space-time manifold as a Weyl integrable
space-time\cite{rom08}.

Let us now restrict ourselves to the particular case when $\omega(\phi
)=\omega=const$. The field equations written above then becomes
\begin{align}
G_{\mu\nu}  &  =-\frac{(\omega-\frac{3}{2})\ }{\Phi^{2}\ }\left(  \Phi_{,\mu
}\Phi_{,\nu}-\frac{g_{\mu\nu}}{2}\Phi_{,\alpha}\Phi^{,\alpha}\right)
\nonumber\\
&  -\frac{1}{\Phi}(\Phi_{,\mu;\nu}-g_{\mu\nu}\square\Phi\ )-\frac{g_{\mu\nu}%
}{2\Phi}V(\Phi)-8\pi T_{\mu\nu},
\end{align}%
\begin{equation}
\square\Phi\ =\frac{1}{\omega}\left(  -\frac{1}{2}\frac{dV}{d\Phi}\Phi
+V(\Phi)\right)  ,
\end{equation}
where we are using the field variable $\Phi=e^{-\phi}$. Note also that we are
expressing the Weylian geometric quantities $\bar{G}_{\mu\nu}$ and
$\bar{\square}\phi$ in terms of their Riemannian counterparts, the latter
being denoted by $G_{\mu\nu}$ and $\square\phi$ calculated from the metric
$g_{\mu\nu}$ and the Christoffel symbols $\{_{\mu\nu}^{\alpha}\}$. \ If we
take $V(\Phi)=2\Lambda\Phi$, which is equivalent to introduce the cosmological
constant $\Lambda$, then the vacuum field equations can be written as%
\begin{align}
G_{\mu\nu}  &  =-\frac{W\ }{\Phi^{2}\ }\left(  \Phi_{,\mu}\Phi_{,\nu}-\frac
{1}{2}g_{\mu\nu}\Phi_{,\alpha}\Phi^{,\alpha}\right) \nonumber\\
&  -\frac{1}{\Phi}(\Phi_{,\mu;\nu}-g_{\mu\nu}\square\Phi\ )-\Lambda g_{\mu\nu
},
\end{align}%
\begin{equation}
\square\Phi\ =\frac{2\Lambda\Phi}{2W+3},
\end{equation}
where $W=\omega-\frac{3}{2}$. On the other hand, the Brans-Dicke vacuum field
equations with a cosmological term $\Lambda$ are given by\cite{kim82}
\begin{align}
G_{\mu\nu}  &  =-\frac{\omega\ }{\Phi^{2}\ }\left(  \Phi_{,\mu}\Phi_{,\nu
}-\frac{1}{2}g_{\mu\nu}\Phi_{,\alpha}\Phi^{,\alpha}\right) \nonumber\\
&  -\frac{1}{\Phi}(\Phi_{,\mu;\nu}-g_{\mu\nu}\square\Phi\ )-\Lambda g_{\mu\nu
},
\end{align}%
\begin{equation}
\square\Phi\ =\frac{2\Lambda\Phi}{2\omega+3},
\end{equation}
which are formally identical to (7) and (8) if we just set $W=\omega-\frac
{3}{2}$. It should be noted, however, that the two theories are not physically
equivalent, since in the Weyl geometrical scalar-tensor theory test particles
follow affine Weyl geodesics (auto-parallels) and not metric
geodesics\cite{rom14}.

\section{Some solutions of the Weyl geometrical scalar-tensor theory}
As a consequence of the formal equality of field equations in the vacuum
regime plus cosmological constant, it is clear that a solution of the Weyl
geometrical scalar-tensor theory can be obtained if we make the change
$\omega\rightarrow\omega-3/2$ in a known solution of Brans-Dicke theory. Thus,
let us consider some solutions of Brans-Dicke theory and obtain the
corresponding solutions in Weyl geometrical scalar-tensor theory.

Initially, let us consider a Friedmann-Robertson-Walker metric with flat
spatial section
\begin{equation}
ds^{2}=dt^{2}-R^{2}(t)[d\chi^{2}+\chi^{2}(d\theta^{2}+\sin^{2}\theta
d\varphi^{2})],
\end{equation}
where $R(t)$ denotes the scale factor. A known class of solutions of the
Brans-Dicke vacuum field equations with a cosmological constant is given
by\cite{bar92}
\begin{equation}
R(t)=R_{0}\exp[(1+\omega)\Psi_{0}t],
\end{equation}%
\begin{equation}
\Phi(t)=\Phi_{0}\exp[\Psi_{0}t],
\end{equation}
where $R_{0}$, $\Phi_{0}$ are constants, and%
\begin{equation}
\Psi_{0}=\pm\sqrt{\frac{2\Lambda}{(2\omega+3)(3\omega+4)}},
\end{equation}
with $\Lambda$ $>0$, $\omega>-\frac{4}{3}$ or $\omega<-\frac{3}{2}$. In this
case, it is not difficult to see that the corresponding solutions of the Weyl
geometrical scalar-tensor theory are simply
\begin{equation}
R(t)=R_{0}\exp\left[  \left(  \omega-\frac{1}{2}\right)  \Psi_{1}t\right]  ,
\end{equation}%
\begin{equation}
\Phi(t)=\Phi_{0}\exp[\Psi_{1}t],
\end{equation}
where, for $\omega>\frac{1}{6}$ or $\omega<0$,%
\begin{equation}
\Psi_{1}=\pm\sqrt{\frac{\Lambda}{\omega(3\omega-\frac{1}{2})}}\,.
\end{equation}

As a second example, let us consider a Friedmann-Robertson-Walker geometry in
the form%
\begin{equation}
ds^{2}=dt^{2}-R^{2}(t)[d\chi^{2}/(1-k\chi^{2})+\chi^{2}(d\theta^{2}+\sin
^{2}\theta d\varphi^{2})].
\end{equation}
For $\ k=1$ and $\omega>-1$, a class of static solutions of the Brans-Dicke
vacuum field equations with $\Lambda$ $>0$ is given by\cite{rom93},%
\begin{equation}
R(t)=\sqrt{\frac{2\omega+3}{\Lambda(\omega+1)}}=const,
\end{equation}%
\begin{equation}
\Phi(t)=\Phi_{0}\exp\left[  \pm\sqrt{\frac{2\Lambda}{2\omega+3}}\,\,t\right]
,
\end{equation}
where $\Phi_{0}$ is a constant. Again, in the Weyl geometrical scalar-tensor
theory we can obtain the following solutions, which are valid for
$\omega>\frac{1}{2}$:
\begin{equation}
R(t)=const=\sqrt{\frac{2\omega}{\Lambda(\omega-\frac{1}{2})}}\,,
\end{equation}%
\begin{equation}
\Phi(t)=\Phi_{0}\exp\left[  \pm\sqrt{\frac{\Lambda}{\omega}}\,\,t\right]  .
\end{equation}

For our purposes, it is also interesting to consider some vacuum solutions
with $\Lambda$ $=0$. Let us now consider the following well known spherically
symmetric solution in Brans-Dicke theory\cite{Bra61}
\begin{equation}
ds^{2}=e^{2\alpha}dt^{2}-e^{2\beta}[d\chi^{2}+\chi^{2}(d\theta^{2}+\sin
^{2}\theta d\varphi^{2})],
\end{equation}%
\begin{equation}
e^{2\alpha}=\left(  \frac{1-B/\chi}{1+B/\chi}\right)  ^{2/\sigma},
\end{equation}%
\begin{equation}
e^{2\beta}=(1+B/\chi)^{4}\left(  \frac{1-B/\chi}{1+B/\chi}\right)  ^{\left(
2/\sigma\right)  \left(  \sigma-C-1\right)  },
\end{equation}%
\begin{equation}
\Phi=\Phi_{0}\left(  \frac{1-B/\chi}{1+B/\chi}\right)  ^{-C/\sigma},
\end{equation}

with,
\begin{align}
\sigma=\left[  (C+1)^{2}-C\left(  1-\frac{1}{2}\omega C\right)  \right]
^{1/2},\nonumber\\
B=\frac{M}{2\Phi_{0}}\left(  \frac{2\omega+4}{2\omega+3}\right)
^{1/2},C=-\frac{1}{\omega+2},
\end{align}
where $\Phi_{0}$ is constant and $M$ is the mass of a spherical matter
distribution. Thus, the analogous solution in the Weyl geometrical
scalar-tensor theory will given by Eqs. (23)-(26), with the new redefined
constants
\begin{align}
\sigma=\left[  (C+1)^{2}-C\left(  1-\frac{1}{2}\omega C+\frac{3}{4}C\right)
\right]  ^{1/2},\text{ }\nonumber\\
B=\frac{M}{2\Phi_{0}}\left(  \frac{2\omega+1}{2\omega}\right)  ^{1/2}%
,C=-\frac{1}{\omega+\frac{1}{2}}.
\end{align}

Another vacuum solution with $\Lambda$ $=0$ is provided by the O'Hanlon-Tupper
solutions\cite{tup72}, with line element given by (11) with
\begin{equation}
R(t)=R_{0}t^{q},
\end{equation}%
\begin{equation}
\Phi(t)=\Phi_{0}t^{r},
\end{equation}
where $R_{0},\Phi_{0},q,r$ \ are constants, $q=\frac{1}{3}(1-r)$ and $\frac
{1}{r}=-\frac{1}{2}\left[  1\pm\sqrt{3(2\omega+3)}\right]  ,$ $\omega
>-\frac{3}{2}$. The corresponding solutions in the Weyl geometrical
scalar-tensor theory are also expressed by Eqs. (29) and (30), with the
redefinition%
\begin{equation}
\frac{1}{r}=-\frac{1}{2}\left[  1\pm\sqrt{6\omega}\right]  ,
\end{equation}
where $\omega>0$.

\section{The Limit $\omega\rightarrow\infty$}
In Weyl geometrical scalar-tensor theory, let us consider a solution of the
scalar field $\Phi$ that takes the form\footnote{Here we would like to point
out that to our knowledge as far as the literature is concerned the precise
meaning of the limit $\omega\rightarrow\infty$ has not been explicitly given.
In fact, the focus has been always on the behaviour of the solutions with
regard solely to the parameter $\omega$. Although not explicitly mentioned,
the procedure seems to take for granted that in the limit $\omega
\rightarrow\infty$ the scalar field $\Phi$ becomes $\Phi=\Phi_{0}+O(1/\omega)$
or $\Phi=\Phi_{0}+O(1/\sqrt{\omega})$ , where both $O(1/\omega)$ and
$O(1/\sqrt{\omega})$ goes to zero for fixed coordinates. More formally, the
meaning of this limit is the following: given an arbitrary $\varepsilon>0$,
there exists a number $N$ such that if $\omega>N,$  then $\left\vert\psi(\omega,x)\right\vert $ $<\varepsilon$ for any fixed values of
the coordinates $x$, where $\psi(\omega,x)=$ $O(1/\omega)$ or $O(1/\sqrt
{\omega})$. In other words, what is required in this procedure is
\textit{pointwise convergence} instead \textit{of uniform convergence.}}%

\begin{equation}
\Phi=\Phi_{0}+O\left(  \frac{1}{\omega}\right)
\end{equation}
for large $\omega$. Now, in the limit $\omega\rightarrow\infty$, we have
$\Phi\rightarrow\Phi_{0}=const$ and the equation (7) may be written as%
\begin{equation}
G_{\mu\nu}+\Lambda g_{\mu\nu}=0.
\end{equation}
Furthermore, as $\Phi=e^{-\phi}$, we get for large $\omega$:%

\begin{equation}
\phi_{,\alpha}=-\frac{\Phi_{,\alpha}}{\Phi}\sim O\left(  \frac{1}{\omega
}\right)  .
\end{equation}
Then, for $\omega\rightarrow\infty$, $\phi_{,\alpha}\rightarrow0$ and the Eqs.
(3) and (4) will be given by%

\begin{equation}
\Gamma_{\mu\nu}^{\alpha}=\{_{\mu\nu}^{\alpha}%
\},\,\,\,\,\,\,\,\,\,\,\triangledown_{\alpha}g_{\mu\nu}=0.
\end{equation}
Let us remark that, in the limit $\omega\rightarrow\infty$, we have a
Riemannian space-time manifold and we recover the Einstein vacuum field
equations with cosmological constant. Moreover, the space-time geometry
becomes Riemannian as can be seen from the behaviour of both the compatibility
conditions and the coefficients of the affine connection.

As a first example, we note that the solution (16) presents the behavior
indicated in Eq. (32) for large $\omega$. Thus, in the limit $\omega
\rightarrow\infty$, we have from (15) that%
\begin{equation}
R(t)=R_{0}\exp\left[  \pm\sqrt{\frac{\Lambda}{3}}t\right]  ,
\end{equation}
becoming identical to de Sitter's solution of general relativity . A second
example, for the particular case $\Lambda$ $=0$, is furnished by the solution
(26), since from Eqs. (28) when $\omega$ is large we obtain
\begin{equation}
\frac{C}{\sigma}\sim O\left(  \frac{1}{\omega}\right)  .
\end{equation}
It is not difficult to see that when $\omega\rightarrow\infty$ the solution
given by Eqs. (23)-(26) and (28) goes over to the Schwarzschild metric of
general relativity, with the identification $1/\Phi_{0}=G$, $G$ denoting the
gravitational constant.

Let us now examine the case in which the solution of the scalar field $\Phi$
for large $\omega$ behaves like%
\begin{equation}
\Phi=\Phi_{0}+O\left(  \frac{1}{\sqrt{\omega}}\right)  .
\end{equation}
In this case, the Eq. (7) in the limit $\omega\rightarrow\infty$ may be
written as%
\begin{equation}
G_{\mu\nu}+\Lambda g_{\mu\nu}=\lim_{\omega\rightarrow\infty}\left[
-\frac{\omega\ }{\Phi^{2}\ }\left(  \Phi_{,\mu}\Phi_{,\nu}-\frac{1}{2}%
g_{\mu\nu}\Phi_{,\alpha}\Phi^{,\alpha}\right)  \right]  \neq0.
\end{equation}
Also, we verify that%
\begin{equation}
\phi_{,\alpha}=-\frac{\Phi_{,\alpha}}{\Phi}\sim O\left(  \frac{1}{\sqrt
{\omega}}\right)
\end{equation}
when $\omega$ is large. It is clear that in this case the Eqs. (3) and (4)
becomes identical to (35) when $\omega\rightarrow\infty$, which then implies
that the space-time manifold becomes Riemannian. However, according to (39),
the Einstein vacuum field equations with cosmological constant (or in the
particular case $\Lambda$ $=0$) are not recovered.

Let us now consider examples of solutions which behave in accordance to (38).
It is immediately seen that the solutions (22) and (30)-(31) satisfy (38) and
that, in these cases, the space-time manifold again becomes Riemannian in the
limit $\omega\rightarrow\infty$. Nevertheless, in the context of the solution
(22), the static solution (21) does not coincide with the general relativity
vacuum solution with $\Lambda$ $\neq0$ when the limit mentioned is taken. The
same conclusion is true in the context of the solution (30)-(31), i.e., the
solution (29) does not represent the Einstein vacuum solution if
$\omega\rightarrow\infty$.

Finally, let us briefly comment on the case in which $T_{\mu\nu}\neq0$ and
$V(\Phi)=0$. We note that if a given solution $\Phi$ satisfies the Eq. (32),
then in the limit $\omega\rightarrow\infty$ the field equations (5) become
\begin{equation}
G_{\mu\nu}=-8\pi T_{\mu\nu}.
\end{equation}
In addition, the Eqs. (35) would also be valid in this limit.

\section{Conclusion}

In this note we have shown that the Weyl geometrical scalar-tensor theory
possesses some similarities with Brans-Dicke theory. Indeed, the two theories
are formally identical if we restrict ourselves to the vacuum field
equations\cite{rom14}. This formal identity also holds if we include the
cosmological constant in the field equations. Due to this identification, we
were able to investigate the limit $\omega\rightarrow\infty$ in the Weyl
geometrical scalar-tensor theory\ simply by getting new solutions from known
solutions of the Brans-Dicke theory. Surely the two theories are not entirely
equivalent, but this simple method allowed us to find solutions in a
straightforward way. We thereby verify that, depending on the behavior of
$\Phi$ for large $\omega$, two possibilities appear when we take the limit
$\omega\rightarrow\infty$. Depending on the asymptotic behaviour of $\Phi$
with respect to large values of $\omega$, i.e., in accordance to Eq. (32) or
to Eq. (38), the field equations reduce or do not reduce to the general
relativistic equations. The space-time geometry, however, becomes Riemannian
in any case.\newline\newline

\textbf{Acknowledgements.} C. Romero thanks CNPq (Brazil) for financial
support. \newline

\end{document}